\documentclass[aps,pre,floats, twocolumn,noshowpacs,superscriptaddress]{revtex4-2}

\usepackage{graphicx,epsfig}
\usepackage{times}
\usepackage{amssymb,amsmath,multirow,rotate,color}

\newcommand{\mean}[1]{\left\langle #1\right\rangle}
\newcommand{\bnu}{{\bm{\nu}}} 
\newcommand{\bmu}{{\bm{\mu}}} 
\newcommand{\blambda}{{\bm{\lambda}}} 
\newcommand{\bq}{{\bm{q}}} 

\usepackage{float}

\begin{document}
\title{Structured interactions as a stabilizing mechanism for competitive ecological communities}
\author{Violeta Calleja-Solanas}
\affiliation{Institute for Cross-Disciplinary Physics and
Complex Systems (IFISC), CSIC-UIB, 07122 Palma de Mallorca,
Spain}
\author{Nagi Khalil}
\affiliation{Complex Systems Group \& GISC, Universidad Rey Juan Carlos, M\'ostoles 28933, Madrid, Spain}

\author{Jes\'us G{\'o}mez-Garde\~{n}es }
\affiliation{GOTHAM Lab., Institute for Biocomputation and
Physics of Complex Systems (BIFI), University of Zaragoza,
50018 Zaragoza, Spain}
\affiliation{Departamento de F{\'i}sica
de la Materia Condensada, Universidad de Zaragoza, 50009
Zaragoza, Spain} \affiliation{Center for Computational Social
Science (CCSS), University of Kobe, Kobe (Japan)}
\author{Emilio Hern\'andez-Garc\'ia}
\affiliation{Institute for Cross-Disciplinary Physics and
Complex Systems (IFISC), CSIC-UIB, 07122 Palma de Mallorca,
Spain}
\author{Sandro Meloni} \email{sandro@ifisc.uib-csic.es}
\affiliation{Institute for Cross-Disciplinary Physics and
Complex Systems (IFISC), CSIC-UIB, 07122 Palma de Mallorca,
Spain}
\date{\today}

\begin{abstract}
How large ecosystems can create and maintain the remarkable biodiversity we see in nature is probably one of the biggest open  questions in science, attracting attention from different fields, from Theoretical Ecology to Mathematics and Physics. In this context, modeling the stable coexistence of species competing for limited resources is a particularly challenging task. From a mathematical point of view, coexistence in competitive dynamics can be achieved when dominance among species forms intransitive loops. However, these relationships usually lead to species'  relative abundances neutrally cycling without converging to a stable equilibrium. Although in recent years several mechanisms have been proposed, models able to explain species coexistence in competitive communities are still limited. Here we identify locality in the interactions as one of the simplest mechanisms leading to stable species coexistence. We consider a simplified ecosystem where individuals of each species lay on a spatial network and interactions are possible only between nodes within a certain distance. Varying such distance allows to interpolate between local and global competition. Our results demonstrate, within the scope of our model, that species coexist reaching a stable equilibrium when two conditions are met: individuals are embedded in space and can only interact with other individuals within a short distance. On the contrary, when one of these ingredients is missing, large oscillations and neutral cycles emerge.

\end{abstract}

\maketitle


\section{\label{sec:intro}Introduction}

The stability of ecosystems is a long-standing question in ecology \cite{May1972,Chesson2018,Allesina2015}. Despite their complexity, ecological systems present remarkable biodiversity that persists for long periods of time. This fact has attracted large attention from several fields in the context of complex systems, in many cases bringing tools from statistical physics or the physics of disordered systems \cite{Bunin2017Ecological,Sidhom2020Ecological}. Throughout the years, multiple mechanisms have been proposed to explain this persistence, including models based on random interactions \cite{May1972} and niche theory \cite{Chesson2018,Bartomeus2018a}. In particular for competitive communities, intransitivity \cite{May1975,Laird2009, kerr2002local} or higher-order interactions \cite{Grilli2017Higher-orderModels,Losapio2019,Levine2017BeyondCommunities,battiston2021physics} have been identified as relevant ingredients to sustain biodiversity. 

Most mathematical models for competitive communities establish a hierarchy among species, where the superior one will drive all the others to extinction, an effect called the competitive exclusion principle \cite{hardin1960competitive}. Despite of it, several mechanisms have been proposed to understand the multiplicity of species observed in natural systems. In particular, the absence of a dominant species can be explained if  dominance among them is established as in a ``Rock-Paper-Scissors'' tournament, where species $i$ out-competes $j$ and $j$ beats $k$, but $k$ is superior to $i$, forming  intransitive cycles. That is, intransitivity may play an important role in the promotion of species coexistence \cite{May1975}, while the structure of the dominance among species may shape their abundance \cite{Laird2009}. Moreover, intransitive tournaments can be defined in probabilistic terms where one species out-competes the other with certain probability; allowing for endogenous stochasticity in the dynamics. 

Concerning stability, the presence of large oscillations in populations is generally considered to be negative for biodiversity maintenance, since species can easily become extinct by external perturbations. Models implementing intransitive dominance often lead species abundances to neutrally cycle around an equilibrium point, something that is unlikely to occur in nature. To overcome this, one of the many approaches that have been proposed is the inclusion of so-called higher-order interactions -- interactions in which the effect of one species on another is modulated by further species \cite{Losapio2019,Levine2017BeyondCommunities}-- leading to convergence to equilibrium, stabilizing the dynamics \cite{Grilli2017Higher-orderModels}. This and other approaches focus on interactions between species and ignore that, within species, single individuals can compete in diverse ways with multiple partners, whose identity can change in time and also in space (\textit{i.e.} ignoring structured interactions).

However, spatial heterogeneity can also have an important impact on species coexistence \cite{valladares2015species,Dieckmann2000,Lowery2019,Travis2005}. The spatial arrangement of individuals can significantly affect the magnitude of their mutual influences, and hence the resulting dynamics. In the same way, the nature of ecological interactions may also shape the spatial distribution of individuals. Diverse works identify space as a driver of coexistence, but it is typically only intended to affect biotic or environmental rates \cite{Travis2005,Dieckmann2000}. The spatial patterns that arise are determined by numerous controlling factors, which can be related to spatial disturbances \cite{Lowery2019}, self-organization processes \cite{Pascual2002ClusterEcologies}, early warning signals of ecological transitions \cite{Kefi2007SpatialEcosystems} or space-dependent ecological interactions \cite{Dieckmann2000}, also with intransitive competition \cite{Laird2009}. Among the ecological processes that depend on spatial location, seed dispersal may have consequences in ecosystem's coexistence and diversity \cite{Liao2016,Chave2013}. However, even if the effect of spatial organization on species coexistence has been in the spotlight for years \cite{kerr2002local}, the question of its role in the emergence and maintenance of stability in competitive intransitive communities, as a way to produce structured interactions, has not been fully explored.

Here, considering  the competitive dynamics that arise from the spatial proximity between sessile individuals,  we demonstrate that space has a stabilizing effect on competitive communities  similar to that induced by higher-order interactions. As a starting point, we study simplified competitive dynamics where competition for resources takes place between pairs of individuals (pairwise interactions) and it is ruled by probabilistic intransitive cycles. We then explicitly introduce space into this framework by defining an interaction network between individuals. Its nodes represent single individuals of different species and links are drawn according to their distance. Positioning individuals in space limits competition to only adjacent neighbors, effectively reducing their mixing. Finally, varying the distance at which links are created allows us to interpolate between local and global interactions and study their effect on the dynamics. This representation provides a suitable context to test whether the spatial distribution of individuals, together with the range of competitive interactions,  may be candidate mechanisms for the maintenance of biodiversity, as alternative to higher-order interactions.

Extensive numerical simulations of our model and of an analytical approximation of the system's dynamics prove that, when we consider only local competition, species abundances naturally converge to the equilibrium without the need of introducing other control mechanisms. These results are built on the fact that there is an underlying spatial structure and are not  attainable by considering interactions of a given individual with just a small number of randomly chosen competitors. On the other side, when the range at which interactions occur increases, abundances start to oscillate in cycles of amplitude increasing with the interaction range. The stabilizing effect of space can be explained by analyzing spatial patterns formed by the species when interactions are local.

In Section \ref{sec:model} we define our model, and describe the results of its numerical simulations in Sect. \ref{sec:results}. We summarize our conclusions in Sect. \ref{sec:Conclusion}. The paper is completed by two Appendices that contain some analytical approaches to the model. 

 \section{\label{sec:model}Competitive Community Model}

We consider an isolated community with a fixed large number of individuals, $N$, each belonging to one of $g$ different species, and model the effect of space in two ways.  Firstly, space affects the arrangement of individuals; which we take into account within a network representation: each individual occupies a node, that symbolizes a fixed spatial location. A node only hosts one individual at a time.  These locations can be regularly spaced or assigned at random. Secondly, two individuals compete if there is a link between them. Links are created according to the interaction range, where short ranges lead to local interactions between nearby nodes. Long-range interactions, instead, result in global competition and loss of spatial correlations.

\begin{figure*}[t!]
 \centering
 \includegraphics[width =0.9\textwidth]{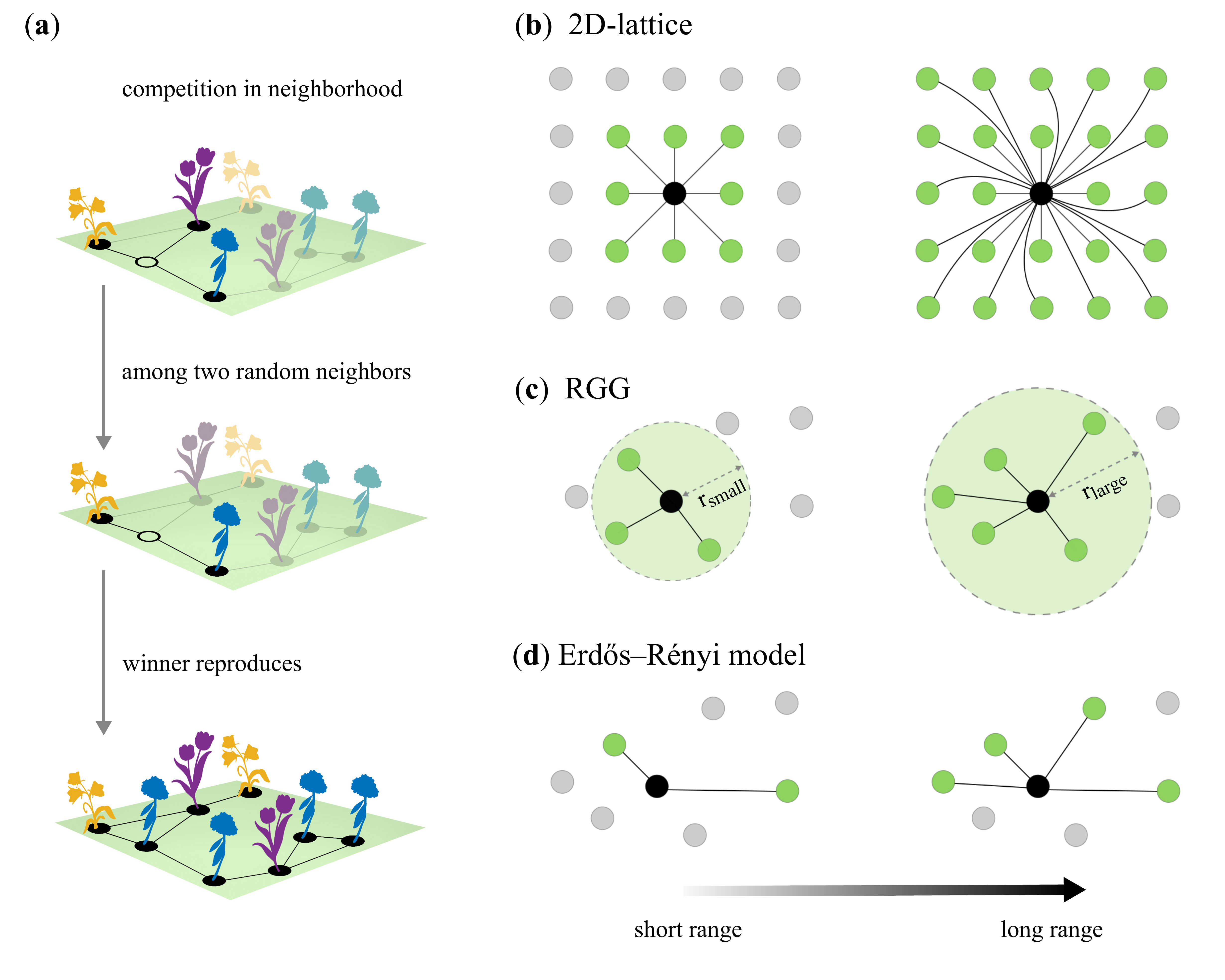}
 \caption{Schematic representation of the interaction networks and competitive dynamics. (Panel a) Diagram of the model. A random plant is selected to die with a probability $1/N$, leaving a vacant fertile region (i.e. an empty node). Two  (highlighted in the middle panel) of the three neighbors are selected at random. Finally, the winner is chosen according to the probabilities of the species dominance matrix $H$, and its descendant sprouts in the vacant node. (Panels b-d) Illustration of the three spatial interaction networks considered. The neighborhood of the black node is depicted (in green) for different interaction ranges. (Panel b) A 2D-lattice with a regular distribution of individuals. The left side of the panel depicts the neighborhood for the
 smallest possible interaction range while the right side highlights the neighborhood when the smallest
 interaction range has been increased by one unit. (Panel c) In a Random
 Geometric Graph, the coordinates of the individuals are uniformly set
 at random in the unit square and two nodes are connected if their Euclidean
 distance is less or equal than $R_{RGG} = r_{small}$ (left side of the panel)
 or $R_{RGG} = r_{large}$ (right side of the panel). (Panel d) Erd{\H{o}}s-R{\'e}nyi
 graphs have no spatial structure. Each pair of nodes connects with probability
 $p$ independently of their distance. The left and right sides of the panel
 illustrate the same  spatial arrangement as in Panel c, but the neighbors
 of the black node are determined  at random by the linking probabilities
 $p = 0.2$ and $p = 0.4$ respectively.
}
 \label{fig:three networks}
\end{figure*}

\subsection{\label{sec:dyn model}Dynamical model}
In order to focus on the interplay between  space and  stability, we keep the number of involved processes to the minimum. Only two ecological processes are present, namely: deaths with identical rate for all species and competition for the vacant location that an individual leaves when it dies. Under these assumptions, our model is suitable for communities of organisms that are permanently attached to one place, as plants. Hence, we describe the model and illustrate our findings through the example of plants competing in a forest. Each plant lives in a fertile region that becomes immediately available after its death.  In that situation, two randomly selected individuals, among all the plants within the interaction range, compete for dispersing their seedlings. This is done via a dominance-matrix approach, as described below. Finally, the winner occupies the vacant node with a descendant of the same species (Figure \ref{fig:three networks}a).

The probability  that a seed of species $i$ wins in a competition with  species $j$, $H_{ij}$, is encoded  in the $g \times g$ dominance matrix $H$. The values of $H_{ij}$ for $i>j$ are drawn  uniformly at random, and we then set  $H_{ji}=1-H_{ij}$, and $H_{ii}  = 0.5$. Within this setting, the system reaches coexistence when $H$ presents intransitive dominance cycles (that occur when $H_{ij} > H_{jk}> H_{ki} > 0.5 $ for some triad $i,j,k$), in accordance with \cite{Grilli2017Higher-orderModels,Allesina2015PredictingWebs}. Specifically, and for sake of reproducibility, in our numerical simulations we employ the following matrix: 
 \begin{equation}
 H = 
 \begin{pmatrix}
0.5 & 0.34 & 0.76 \\
0.66 & 0.5 & 0.25 \\
0.24 & 0.75 & 0.5 
\end{pmatrix}. \label{eq:H}
 \end{equation}
Moreover, given the form of $H$, the ecosystem is constrained in the long-term to have an odd number of species \cite{Grilli2017Higher-orderModels}. When one species vanishes, another extinction event must occur to maintain the odd number of species.

\subsection{Interactions' structure}
To explore the effect of spatial arrangement, we employ three different types of networks: a 2D square lattice, a Random Geometric Graph \cite{Dall2002RandomGraphs} and an Erd{\H{o}}s-R{\'e}nyi graph \cite{erdos1959random}. Each network defines a certain type of spatial distribution.

A \textit{2D square lattice} is our baseline for a highly-ordered space because of its simplicity and wide use in ecology \cite{Dieckmann2000,grimm2005IBM,Lowery2019}. Nodes are regularly distributed on the unit square and are at a discrete, constant distance apart from each other.  The nearest neighbors of a node are considered to be the eight adjacent nodes (with periodic boundary conditions) (Figure~\ref{fig:three networks}b). This network, since nodes are regularly spaced and connected, can generate strong spatial correlations.

In addition to lattices, we consider a \textit{Random Geometric Graph} (RGG) that conserves the spatial structure but in a disordered  manner, as  the $N$  nodes are  uniformly distributed in the unit square and two of them are linked if their Euclidean distance is smaller or equal to a particular interaction range $R_{RGG}$ (Figure~\ref{fig:three networks}c) allowing us to study continuous distances and variability in the number of neighbors \cite{cardillo2012,estrada2016,arias2018}.

Finally, we consider non-spatial interactions through \textit{Erd{\H{o}}s-R{\'e}nyi graphs} (ER), where nodes are connected at random with probability $p$ and, hence, the location of individuals does not affect their linking probability (Figure~\ref{fig:three networks}d). In this case, spatial correlations are completely destroyed, although each node still has a finite number of neighbors.

Summing up, the ER graph is our null model since it has no spatial structure, while we include the RGG as a compromise between unstructured and regularly-spaced interactions.

We tune the competition from local to global in the different networks by means of the interaction range. This range determines the individuals that participate in the competition, \textit{i.e.} who interacts with whom. With short-range interactions, only nearby nodes compete. As it increases, more distant nodes enter the competition until the neighborhood size is large enough to dissolve the effect of location and consider the system well-mixed. In particular, for square lattices, this leads to connections between not only the closest nodes but also the second, the third groups of neighbors, etc. Meanwhile, increasing the interaction range in a RGG means increasing the distance $R_{RGG}$. Finally, position or distances between nodes do not enter into the construction of ER networks. In this case,  the connection probability $p$ serves as a proxy for the interaction range. Increasing $p$ generates larger neighborhoods, albeit their location is at random. In order to use a quantity that can be compared with the other networks, it is convenient to quantify  the interaction range by the mean degree $\langle k \rangle =pN$. For every network, we trivially get all-to-all competition with the largest interaction range.

\begin{figure*}[t!]
     \centering
\includegraphics[width=0.92\textwidth]{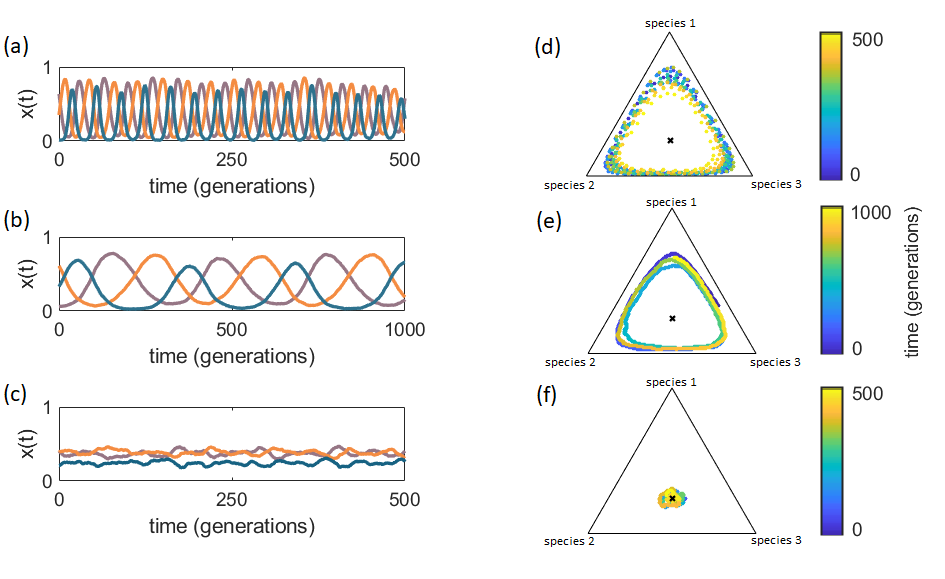}
 \caption{(Panels a,b,c) Temporal evolution of the species relative abundances for a 3-species system, comparing different interaction schemes.
  In each panel, the ecosystem  is represented by a RGG of $ N = 10^4$ nodes.
  Relative abundances $x_1,x_2,x_3$ are plotted after the transient has vanished. (Panel a) All-to-all interactions: the range is set to cover the entire plane ($R_{RRG} = R_{max} = \sqrt{2}$), hence individuals can compete for any vacant node. (Panel b)  Long-range interactions: we set $R_{RGG} = 0.15$ leading to an average degree $\langle k \rangle \simeq 706$. (Panel c) Short-range interaction: $R_{RGG} = 0.03$ and $\langle k \rangle \simeq 28$. (Panels d,e,f) Trajectories in the phase space represented on the standard 2-simplex (the portion of the $x_1 + x_2+ x_3 = 1$ plane in which $x_1,x_2,x_3 \geq 0$).  The plots show a view perpendicular to the simplex, and correspond to the time evolution of the left panels. The color code represents time evolution. With all-to-all and long-range interactions (Panels d and e), abundances  oscillate in large cycles around what seems to be an  equilibrium point (represented by a black cross). With short-range interactions (Panel f), abundances remain confined in a small region around the equilibrium.}
\label{fig:three graphs}
\end{figure*}

\section{\label{sec:results}Results}
Once our model has been defined, we start analyzing it through extensive Monte Carlo simulations. At the beginning of each simulation, the species within each node is assigned at random with a uniform probability $1/g$. We simulate the system using an asynchronous update scheme, where a generation is defined as $N$ updates to ensure that, on average, every node has experienced a death event. Finally, we keep track of the proportion  or relative abundance of individuals of each species in the system, $x_i(t) \equiv N^{-1}\sum_\nu^N n_{i,\nu}$, where $n_{i,\nu}$ takes the value 1 if and only if species $i$ is present at node $\nu$. Each node can host only one individual of a single species, which implies that $\sum_i^g n_{i,\nu}=1$, $\forall \nu$. Since the total number of nodes in the system is constant and equal to the total number of individuals $N$,  the macroscopic quantities $x_i$ are also average total spatial densities.

Since we have $\sum_{i}^{g} x_i(t) = 1$ for every generation $t$, the relative abundances of all species can be represented by a point in the $(g-1)$-simplex $\{(x_1,...,x_g) | x_i \geq 0 \textrm{ and } \sum_{i=1}^{g} x_i=1\}$, whose vertices correspond to single-species populations. As time evolves, the point follows a trajectory on the simplex that characterizes the macroscopic state of the system.

\subsection{\label{sec:temporal evo}Temporal evolution}

We begin our analysis by inspecting the temporal evolution of species' abundances in the simplest situation of three competing species, $g=3$. Unless otherwise stated, we use always the same matrix $H$ given in Eq.~(\ref{eq:H}), which gives results representative of any other randomly generated dominance matrix with intransitive cycles. We find different behaviors depending on the spatial distribution of species and the distance at which they interact. Species in communities with no spatial structure (all-to-all interactions, Figure~\ref{fig:three graphs}a; same result for ER graphs) cycle on the simplex. The same wide oscillations (Figure~\ref{fig:three graphs}b) can also be seen if we consider long-range interactions in structured communities (RGG and 2D-lattice). This first result is in line with the prediction of the mean-field approximation (see Appendix \ref{sec:theory}). However, the amplitude of the observed oscillations is independent of the initial conditions, indicating that these oscillations are of the limit-cycle type, qualitatively different from the neutral ones predicted by the mean-field theory.

For the two spatial networks considered, decreasing the interaction range leads to a reduction in the amplitude of the oscillations until, for a sufficiently short-range, species' abundances only slightly fluctuate around an equilibrium state (Figure~\ref{fig:three graphs}c).

Their value at this point is, in all cases analyzed, close to the equilibrium fixed point obtained from the mean-field approximation (which for the matrix $H$ in Eq.~(\ref{eq:H}) is $(x_1,x_2,x_3)=(0.374, 0.383, 0.243)$). These values also coincide with the temporal average of the relative abundances in the oscillatory case for the same matrix $H$.

These latter results reveal a non-trivial dependency of the dynamics on the interaction range, and demand a deeper analysis. For this purpose, in the next sub-sections, we systematically study the effect of the interaction range and structure on species' dynamics.

\subsection{\label{sec:Phase space} Dynamical behavior depends on structured interactions} 

As a first step, we need a measure to characterize the behavior of the system for each structure and interaction range. Because of the noisy character of the dynamics in the stochastic simulations, the amplitude of the oscillations is not a robust indicator. Instead, we consider the area encircled by the system's trajectory on the simplex. If the system fluctuates with small amplitude around some equilibrium abundances, the trajectory occupies a small area (Figure~\ref{fig:three graphs}f), whereas larger oscillations would cover broader areas (Figure~\ref{fig:three graphs}d,e).

\begin{figure}[t!]
    \centering
    \includegraphics[width=0.51\textwidth]{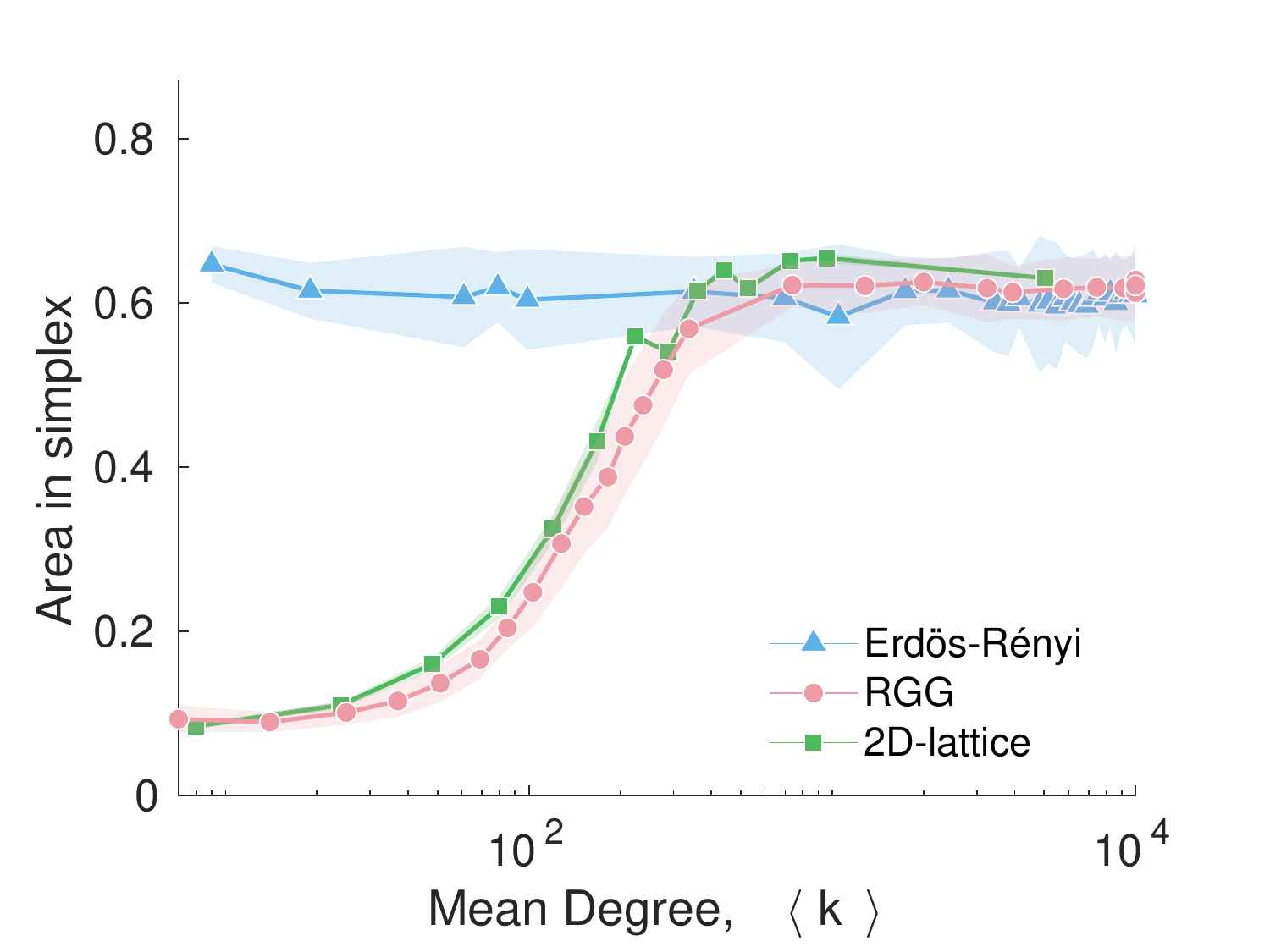}
    \caption{Average area inside the trajectory on  the 2-simplex of the $(x_1,x_2,x_3)$ point of a 3-species community (see Figure~\ref{fig:three graphs}, Panels d-f) as a  function of average degree $\langle k \rangle$, for different networks. The total number of individuals is $N = 10^4$, and the same dominance matrix $H$ is used for all networks. The points represent the mean area obtained over $50$ realizations, each simulated in different networks. Areas have been calculated excluding the $5\%$ of out-layer points in the trajectory. Shaded areas indicate $95\%$ confidence interval.}
    \label{fig:areasPaper}
\end{figure}

 Once defined our metric to characterize the stability of the dynamics, we can study the effect of space by keeping $H$ fixed in all the simulations and varying the underlying network structure (the type of graph) and the interaction range. Since we cannot properly define distances in ER graphs, we use the degree as a proxy of interaction range for that graph. This equivalence can be made as the interaction range not only defines the distance at which nodes compete but also their degree. In that way, we are ready to compare the two spatial networks with the ER graphs. 

To start with, we focus on the effect of the interaction network but without any spatial arrangement  by  considering the ER graphs with increasing average degree: \textit{i.e.} increasing $p$ (blue points in Figure~\ref{fig:areasPaper}). We find that the dynamics show large oscillations for all values of the degree. That is, the size of the neighborhood does not affect the dynamics.

However, this picture drastically changes when we consider spatially structured interactions.

We recover the results of the ER networks for large ranges (large average degrees) in both the RGG and the 2D lattice. However, the system stabilizes around the equilibrium point when we decrease the interaction range, covering a tiny area in the phase space. The transition between these two regimes takes place when the average degree of both networks is within the range $ 50 \lesssim \langle k \rangle \lesssim 100$.

\begin{figure*}[t!]
 \centering
  \includegraphics[width=1.0\linewidth]{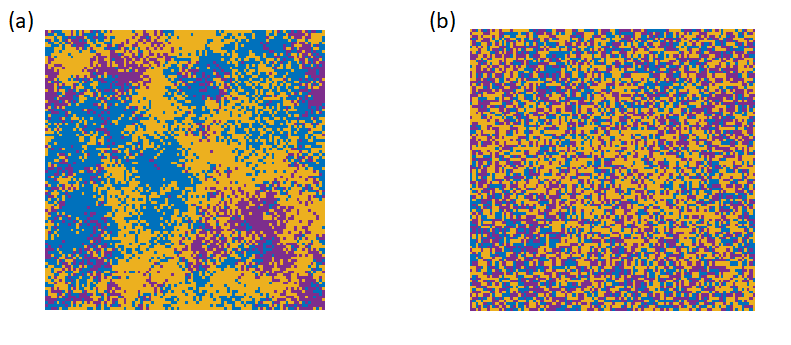}
\caption{Spatial organization of a 3-species community in a 2D-lattice of $N=10^4$ for short and long-range interactions with the dominance matrix given in Eq.~(\ref{eq:H}). Individuals of each species are depicted in a different color. (Panel a) Short-range interactions: when a plant dies only the $8$ closest neighbors at distance one compete for a vacant node (see the left network from Figure~\ref{fig:three networks}b). (Panel b) Long-range interactions: the $360$ individuals at distance less or equal to $9$ from a vacant node participate in the competition.} \label{fig:ecosystems}
\end{figure*}

To summarize, the intuitive picture that arises from these results is the following: when we consider long-range interactions -- e.g. large degrees --  we obtain large oscillations, which are similar to the ones obtained for non-spatial networks (ER). In all cases, the amplitude and period of the oscillations are independent of the initial conditions, i.e. the oscillations are of the limit-cycle type. The mean-field approximation (see Appendix \ref{sec:theory}), which is expected to be valid in the limit of long-range interaction, correctly predicts oscillatory behavior. But it fails to reproduce the limit-cycle character, predicting neutral oscillations instead.  When we restrict competition to small neighborhoods  (small degrees) we find that the dynamics stabilizes around some fixed point $x^*$. 

\subsection{\label{sec:Spatial} Spatial configurations}

So far we have only considered the trends of the global relative abundances, $x_i$, quantities that are influenced by, but do not explicitly display information on, the spatial distribution of individuals. To better understand the mechanism behind the reported behavior, we show in Figure~\ref{fig:ecosystems} two different snapshots of the spatial organization of a 3-species system  in a 2D square lattice for two different interaction ranges (short and long).

With short ranges (Figure~\ref{fig:ecosystems}a), species self-organize in mono-specific patches. Changes in species relative abundances can only take place along the borders, where different species meet. A death event inside the patch does not contribute to relative abundance variations because competition is among same-species individuals. In this way, patches are more robust to invasion from other species, decelerating the dynamics of the system and hence the possibility of heavy oscillations.

Differently, with long-range interactions (Figure~\ref{fig:ecosystems}b), the unstructured and statistically homogeneous solution predicted by  mean-field theory appears: vacant nodes can be reached by any species blocking the formation of single-species clusters. The absence of patches prevents the community from reaching a steady state, with intransitive cycles generating large-scale oscillations.

Taken together, these latter results suggest that short-range interactions reduce the effective competition in the system by decreasing the probability of an encounter between individuals of different species. To confirm this hypothesis, we calculate the average probability $\langle P_{ij} \rangle$ that two species $i$ and $j$ compete for a vacant node in the short-range regime and compare it with the expected value $\overline{P}_{ij}$ in the all-to-all case. $\langle P_{ij}\rangle$ has been obtained numerically by recording the number of times species $i$ and $j$ have been selected for competition and then averaging over the duration of the simulation. For all-to-all interactions, $\overline{P}_{ij}$ is given by the product of the relative abundances of species $i$ and $j$ at the mean-field equilibrium abundances $\overline{P}_{ij}=x_i^* x_j^*$ (see Appendix \ref{sec:theory}). For our example system we have $x^*=(0.374, 0.383, 0.243)$, so
that:
 \begin{equation}
\overline{P}_{ij} =
\begin{pmatrix}
0.1399 & 0.1432 & 0.0909\\
0.1432 & 0.1467 & 0.0931\\
0.0909 & 0.0931 & 0.0590
\end{pmatrix} 
\label{eq:mata2a}
\end{equation}
The computation of matrix $\langle P_{ij} \rangle$ for Eq.~(\ref{eq:H}), in a RGG with short-range interactions ($R_{RGG} = 0.022$ and $\langle k \rangle \simeq 15$) gives the following result:
\begin{equation}
\langle P_{ij} \rangle =
\begin{pmatrix}
\textbf{ 0.2160} & 0.0965 & 0.0597\\
 0.0965 & \textbf{0.2241} & 0.0659\\
 0.0597 & 0.0659  & \textbf{ 0.1156}
\end{pmatrix} \ .
\label{eq:matshort}
\end{equation}
We see that, when compared to the all-to-all case, for short-range interactions, same-species competition has a higher probability to occur ($\langle P_{ii} \rangle$, highlighted in boldface in Eq. (\ref{eq:matshort})) than different-species competition (the off-diagonal terms). This demonstrates that spatial inhomogeneities reduce the effective inter-specific competition. Finally, as a further confirmation of this mechanism, in Appendix \ref{sec:app_toymodel} we show that a toy model, based on the mean-field formulation of the model but where inter-specific interactions are reduced and intra-specific ones are increased, presents the same shift in stability observed in our spatial models.

\begin{figure*}[t!]
 \centering
 \includegraphics[width=0.99\textwidth]{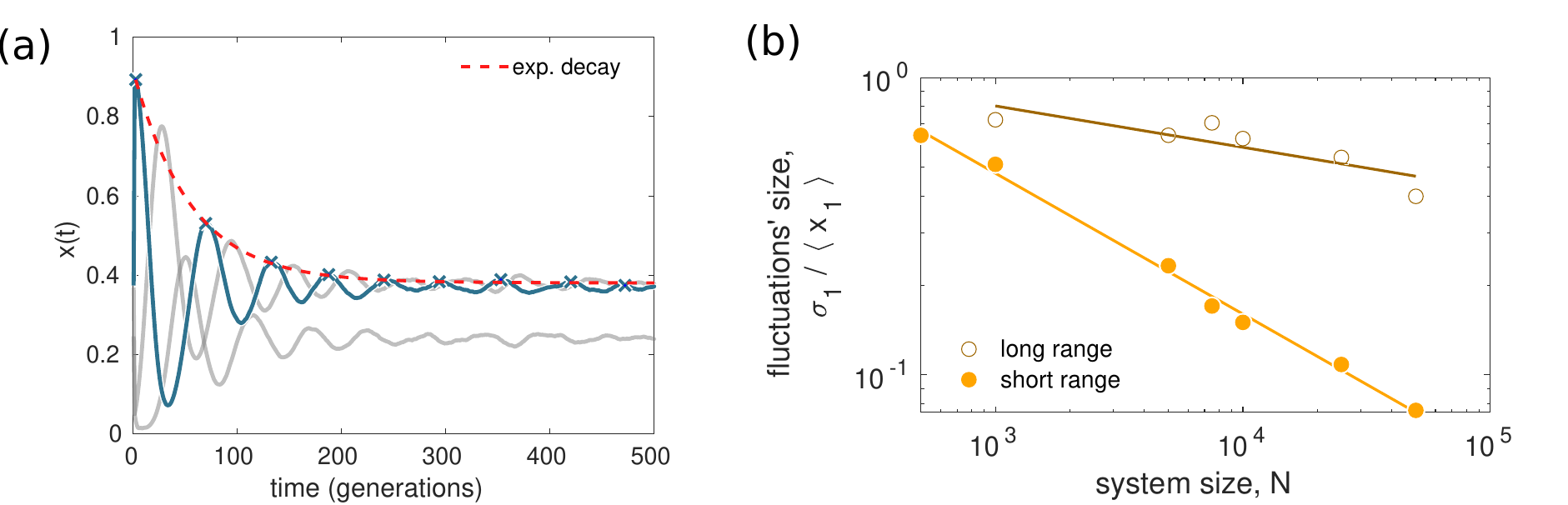}
  \caption{(Panel a) Time evolution of the recovery from a $90 \%$ pulse perturbation in a 3-species community for the dominance matrix $H$ of Eq.~(\ref{eq:H}). The relative abundance  of one species (blue) is artificially modified from its equilibrium value to be the $90 \%$ of the whole population, whereas  other species' relative abundances (in grey) are proportionally decreased. The simulation is performed in a RGG of $10^4$ individuals and $R_{RGG} = 0.03$. The red line represents the fit of the local maxima of the relative abundance (blue crosses) to the function $a e^{-\alpha} + b$ with $\alpha = 0.018$, $a = 0.53$ and $b = 0.38$. (Panel b) Scaling of fluctuations, measured by the coefficient of variation of species 1 ($\sigma_1/\mean{x_1}$) with the system size $N$ for  a RGG with $3$ species. Each point is the result of $10$ different realizations where the variance and mean relative abundance of species 1 have been calculated over at least $\Delta t = 10^8$ time-steps after the transient. Short-range interactions correspond to an average degree $\langle k \rangle = 15 \pm 2$, and we find a decrease of the relative fluctuations with system size as $\sigma_1/\mean{x_1}\sim N^{-0.47}$, consistent with an scenario of uncorrelated domains. For a situation of long-range interactions we set $\langle k \rangle = 980 \pm 190$, giving a scaling of the relative fluctuations as $N^{-0.14}$.}
    \label{fig:scaling}
\end{figure*}

\subsection{\label{sec:scaling}Stability and fluctuations}
Once clarified the mechanism behind the stabilization of the dynamics for short interaction ranges, we conclude our analysis by probing further the stability of the fixed point for the
macroscopic variables $x_i$, and by studying the nature of the fluctuations around it that are seen in the  simulations. 

To check the stability of the equilibrium reached, we study the system's response to pulse perturbations of different magnitudes. In our model, this translates into imposing a sudden change in species' relative abundances and measuring the time needed to recover the original state. Figure~\ref{fig:scaling}a shows the results for a RGG  for $R_{RGG}=0.03$ (short range), with a $90 \%$ perturbation of one species' relative abundance (with all other abundances being proportionally decreased), demonstrating that, even with such a large disruption, the dynamics bounce back to the equilibrium as
the perturbation decays exponentially in time.

Finally, we study the characteristics of the fluctuations around the equilibrium for both the stable and the unstable regimes. To do so, we focus on how the size of fluctuations in the relative abundance of each species (defined as their coefficient of variation, $\sigma_i/\langle x_i \rangle$) scales with the size of the system. In Figure~\ref{fig:scaling}b we show the scaling for one species in a RGG. For small degrees ($\langle k \rangle = 15 \pm 2$), we find an exponent of $0.47$, pretty close to the $0.5$ expected in case of  residual fluctuations arising from many nearly uncorrelated domains and the stochastic noise due to the finite size of the system. This rules out the possibility that the observed fluctuations originate from the presence of oscillatory behavior of small amplitude. In turn, for the unstable case (large degrees, $\langle k \rangle = 980 \pm 190$) we observe an exponent of $0.14$. In this case, fluctuations are a genuine ecological signal that emerges from the interactions in a high-mixing environment.

\section{Discussion and conclusions}
 \label{sec:Conclusion}
 
Many efforts have been made to explain the remarkable robustness observed by natural ecosystems in terms of biodiversity.
These efforts include niche and neutral models and higher-order interactions. Here, considering a minimal model for competitive communities, we have proved that spatial interactions alone lead to the coexistence and stability of multi-species systems.

In particular, making use of extensive numerical simulations we have studied a simple model where multiple species compete in a structured space in intransitive dominance cycles. Analyzing different spatial arrangements, ranging from regular lattices to random connections that cancel out the effect of space, our results show that spatial interactions limited to nearest neighbors lead to stable coexistence of different species, while for long-range interactions species' relative abundances indefinitely oscillate. By taking into account the spatial organization of the individuals, we discovered that local interactions allow species to survive by forming mono-specific patches where competition only takes place at their borders and, as result, decreasing the effective competition experienced by each individual. This latter effect generates a deceleration of the dynamics, effectively damping out fluctuations. These last results, however, are not matched by mean-field approximations, as described in Appendix \ref{sec:theory}. This is not surprising since the dynamics depend strongly on the nature of the spatial correlations created by the finite-range interactions.

In conclusion, even if our results are obtained with a simplified model, taken together our findings help to explain the role of space in maintaining stable spatial coexistence in natural ecosystems. In this sense, a restricted interaction range goes against the coherent and neutral oscillatory behavior usually produced by intransitive interactions. While in real ecosystems many simultaneous mechanisms may be at play, as for example higher-order interactions, spatial effects are probably the simplest and most widely present of them, and thus they need to be considered when addressing ecological coexistence.

\acknowledgments V.C-S. and S.M. thank Hugo Saiz for insightful discussions. J.G.G. acknowledges support by Gobierno de Aragon/Fondo Social Europeo (E$36\_20$R FENOL group), grant FIS2017-87519-P funded by Spanish MINECO, and grant PID2020-113582GB-I00 funded by MCIN/AEI/10.13039/501100011033.E.H-G., S.M. and V.C-S. acknowledge the Spanish State Research Agency, through the Severo Ochoa and Mar\'ia de Maeztu Program for Centers and Units of Excellence in R\&D (MDM-2017-0711) funded by MCIN/AEI/10.13039/501100011033. V.C-S. acknowledges support from FPI/2257/2019 Conv. 201, CAIB PhD program. N.K. acknowledges support by Rey Juan Carlos University (Grant M2605) and by Community of Madrid and Rey Juan Carlos University through Young Researchers program in R\&D (Grant CCASSE M2737).

\appendix

\section{\label{sec:theory}Analytical formulation}

Along with a numerical implementation of the dynamics, it is
also possible to provide a mathematical description of the
model, which we set up in this Appendix. In Section
\ref{sec:moments} we establish the basic equations for the
moments of the population variables. Sect. \ref{sec:mf}
develops a standard mean-field approximation for statistically
homogeneous systems. We stress that it is not able to reproduce
the main numerical findings for our model, but gives a baseline
to interpret the results. Sect. \ref{sec:localmf} extends the
mean-field approximation to allow for spatial inhomogeneity in
the species distribution. The results still do not match with
the numerical observations, but give some hints on the reduced
stability of homogeneous oscillations when the interaction
range is small.

\subsection{Moment equations}
\label{sec:moments}

An analytical description of the stochastic dynamics defined in
the main text can be given (after a trivial replacement of the
discrete-time dynamics by a continuous-time one) by the master
equation for the time-dependent probability of the system
state. It allows us to derive equations for the expected
relative abundance of each species at a given node as well as
for the two-node correlations.

The model state can be specified by giving $\{Z_\nu\}$, where
$Z_\nu=1,2,...,g$ specifies the species that occupies node
$\nu\in\Sigma$, with $\Sigma$ being the set of nodes of the
network. However, we find more convenient to parameterize the
model as follows. Let $n_{i,\nu}\in \{0,1\}$ be the number of
individuals of species $i\in\{1,\dots,g\}$ at node $\nu\in
\Sigma$, i.e. $n_{i,\nu}=1$ for one and only one $i$,
identifying the species present at $\nu$, and $0$ for the other
values of $i$ (absent species). The state of the system can be
characterized by the set of vectors $S=\{S_\nu\}_{\nu=1}^N$,
with $S_\nu=\{n_{1,\nu},\dots,n_{g,\nu}\}$. This state evolves
as follows: (i) with a rate $r$, a randomly chosen individual
(say, located at $\nu$) dies, then (ii) two neighbors of the
dead individual (thus pertaining to the set $P_\nu$ of
neighbors of $\nu$) are chosen at random and compete to
generate the offspring: a winner species is selected according
to the probabilities in the dominance matrix $H$. And (iii)
this offspring is immediately located at the vacant node.
Following standard procedures (for example see
 \cite{khlohe17,klkh20}) the master equation for the probability
$p(S,t)$ of finding the system in a state $S$ at time $t$ can
be written as
\begin{align}
  \label{eq:me}
  \frac{\partial}{\partial t}p(S,t)=
  \sum_{\nu=1}^N\sum_{i,j}\left( E^+_{i,\nu}E^-_{j,\nu}-1\right)\pi_\nu(i\to j)p(S,t),
\end{align}
where the operators $E^\pm$ act on an arbitrary state function
$f(S)$ as
\begin{eqnarray}
  \nonumber
E^\pm_{i,\nu}f(S)=f&\Big(&\{n_{1,1},\dots,n_{g,1}\} ,\dots, \\   \nonumber
&&\{n_{1,\nu},  \dots,n_{i,\nu}\pm 1,\dots,n_{g,\nu}\},\dots,\\
&&  \{n_{1,N},\dots,n_{g,N}\} \ \ \Big).
\end{eqnarray}
$\pi_{\nu}(i\to j)$ is the rate at which an individual of
species $i$ is replaced by one of species $j$ at site $\nu$,
given by
\begin{equation}
  \pi_{\nu}(i\to j)=r\frac{n_{i,\nu}}{N}\frac{2}{k_\nu(k_\nu-1)}
  \sum_{\stackrel{\lambda,\mu\in P_\nu}{\mu\neq \lambda}}\sum_{k}n_{j,\lambda}n_{k,\mu}H_{jk},
\end{equation}
where $k_\nu$ is the degree of node $\nu$, i.e. the number of
nodes in $P_\nu$.

From the master equation we can derive equations for the
moments of the distribution, which can be easily measured from
the numerical simulations. The simplest nontrivial moment is
the expected number of individuals of species $i$ at node
$\nu$, $\mean{n_{i,\nu}}$. Its equation is readily obtained
from the master equation after multiplying it by $n_{i,\nu}$
and summing over all possible values of $S$:
\begin{eqnarray}
  \label{eq:ni}
  \frac{d}{ds} \mean{n_{i,\nu}} =&& \frac{1}{k_\nu(k_\nu-1)}
  \sum_{j} \sum_{\stackrel{\lambda,\mu\in P_\nu}{\mu\neq \lambda}}
  H_{ij} \mean{n_{i,\lambda}n_{j,\mu}}\nonumber \\
  && - \frac12 \mean{n_{i,\nu}} ,
\end{eqnarray}
where we have introduced a new time scale $s\equiv
\frac{2r}{N}t$.

From this equation we can write the dynamics for the expected
value of the macroscopic variable $x_i(s) \equiv N^{-1}\sum_\nu
n_{i,\nu}$ as 
\begin{equation}
\label{eq:xintermsofP}
  \frac{d}{ds} \mean{x_i(s)} =
  \sum_{j} H_{ij} P_{ij}(s) - \frac12 \mean{x_i(s)} ,
\end{equation}
where we have introduced the symmetric matrix
\begin{equation}
\label{eq:Ps}
  P_{ij}(s) = \frac{1}{N} \sum_\nu \frac{1}{k_\nu (k_\nu -1)}
  \sum_{\stackrel{\lambda,\mu\in P_\nu}{\mu\neq \lambda}} \mean{n_{i,\lambda}n_{j,\mu}} \ .
\end{equation}
This matrix can be interpreted as the probability of sampling
at time $s$ a pair of individuals of species $i$ and $j$ when
deciding the replacement of a dead individual somewhere in the
system. It satisfies  $\sum_{ij} P_{ij}(s) =1$
and, in a homogeneous network ($k_\nu=k$, $\forall \nu$),
$\sum_{j=1}^g P_{ij}(s)=\mean{x_i(s)}$.

As for the second-order moments, both for $\mu \in P_{\nu}$ and
for $\mu\notin P_\nu$, their equations read
\begin{eqnarray}
\nonumber
  \frac{d}{ds} \mean{n_{i,\nu}n_{j,\mu}} =&&\frac{1}{k_\nu(k_\nu-1)}
  \sum_{l}\sum_{\stackrel{\delta,\lambda\in P_\nu}{\delta\ne \lambda}}H_{il}\mean{n_{i,\lambda}n_{l,\delta}n_{j,\mu}} \\
  && + \frac{1}{k_\mu(k_\mu-1)}
  \sum_{l} \sum_{\stackrel{\delta,\lambda\in P_\nu}{\delta\ne \lambda}}H_{jl}\mean{n_{j,\lambda}n_{l,\delta}n_{i,\nu}} \nonumber \\ &&-\mean{n_{i,\nu}n_{j,\mu}}.
\label{eq:ninj}
\end{eqnarray}

In general, it can be seen that the moment equations form a
hierarchy, namely that the equation for a moment of order $o$
depends on the moments of order $o+1$. Hence, they cannot be
solved in closed form, except if introducing some
approximation.

\subsection{Homogeneous mean-field approximation}
\label{sec:mf}

The simplest of such approximations is the mean-field approach.
It is conveniently done in the simplified case in which the
network is spatially homogeneous, i.e. all the nodes have the
same degree: $k_\nu=k$, $\forall \nu$. In this situation, we
can search for statistically homogeneous solutions:
$\mean{n_{i,\nu}(s)}=\rho_i(s)$, $\forall \nu$. We can relate
these time-dependent moments $\rho_i(s)$ to the macroscopic
variables $x_i(s) \equiv N^{-1}\sum_\nu n_{i,\nu}(s)$ (note
that $\sum_{i=1}^g x_i(s)=1$). Indeed we have
$\mean{x_i}=N^{-1}\sum_\nu \rho_i=\rho_i$, or
$\mean{x_i}=\mean{n_{i,\nu}}$.

The mean-field approximation, which is exact in the case of
all-to-all interactions in an infinite system, and expected to
be accurate both for large enough interaction range (mean
degree) and for unstructured interactions, consists in
neglecting fluctuations and correlations:
\begin{eqnarray}
& \mean{n_{i,\nu}}=\mean{x_i}  \simeq x_i\ ,  & \forall \nu\in \Sigma, \\
& \mean{n_{i,\nu} n_{j,\mu}}   \simeq \mean{n_{i,\nu}} \mean{n_{j,\mu}} \simeq x_ix_j \ , &  \forall \nu\ne \mu \in \Sigma.
\end{eqnarray}
We have also $P_{ij}\approx x_i x_j$. Introduction of these
expressions into Eq. (\ref{eq:ni}) leads to a closed evolution
equation for $x_i$:
\begin{equation}
  \label{eq:meanf}
  \frac{d}{ds} x_i = \left(\sum_{l}H_{ij}x_j-\frac12\right)x_i \ .
\end{equation}

This mean-field equation has been studied before (e.g.
 \cite{Grilli2017Higher-orderModels}). We summarize here the
main results:

First, the dynamics (\ref{eq:meanf}) maintains in time the
property $\sum_i x_i =1$, if the initial condition satisfies
it. This can be seen by defining $X\equiv \sum_i x_i$,
calculating $dX/ds$, using that $H_{ij}=(H_{ij}+H_{ij})/2 =
(1-H_{ji}+H_{ij})/2$, and noticing that $\sum_{ij}
(H_{ij}-H_{ji})x_i x_j =0$ and $\sum_{ij} x_i x_j =X^2$. Thus
the sum of relative abundances satisfies
\begin{equation}
\label{eq:sumdynamics}
\frac{dX}{ds} = \frac12 (X^2 - X) \ ,
\end{equation}
which maintains $X(t)=1$, $\forall t$ if
$X(0)=1$.

Second, Eq. (\ref{eq:meanf}) has several equilibria or fixed
points. Many of them are of the `absorbing' or `boundary' type,
i.e. steady solutions of (\ref{eq:meanf}) in which $x_i=0$ for
some $i$, so that the corresponding species are extinct. In
addition, if $g$ is odd, there is generically
 \cite{Grilli2017Higher-orderModels} an \textit{interior}
equilibrium, $x_i(t)=x_i^*$, $\forall t$, in which all species
coexist with non-vanishing relative abundances $x_i^*$. At this
fixed point the relative abundances are given by
\begin{equation}
  \label{eq:sts}
  \sum_{j=1}^g H_{ij} x_j^* =\frac12 \ \ \Rightarrow \ \
  x_i^*=\frac12 \sum_j (H^{-1})_{ij},
\end{equation}
where $H^{-1}$ is the inverse of the dominance matrix, which
always exists when it describes an intransitive loop. The
properties of the boundary fixed points can be analyzed by
recognizing that they can be considered interior equilibria in
a system with a smaller number $g$ of species.

Third, the dynamics from arbitrary initial conditions in which
all $x_i$ are non-vanishing (and for generic $H$) leads to a
transient in which some of the species may become extinct. The
remaining ones, an odd number, cycle neutrally around the
interior fixed point (\ref{eq:sts}) in which the rows and
columns corresponding to the extinct species have been removed
from $H$ \cite{Grilli2017Higher-orderModels}. The stability of
this interior equilibrium is always neutral: relative
abundances of surviving species describe periodic closed orbits
around it, with an amplitude and period that is determined by
the initial condition and without being attracted nor repelled
by the fixed point. This can be seen
 \cite{Grilli2017Higher-orderModels} by noticing that the
quantity
\begin{equation}
V(x_1,...,x_g) = - \sum_{i=1}^g x_i^* \log \frac{x_i}{x_i^*}
\end{equation}
is a constant of motion, and thus it foliates the
$(g-1)$-simplex on which the dynamics occurs into invariant
hypersurfaces that turn out to contain concentric closed orbits
around the interior equilibrium.

The neutral character of the oscillations is not realistic from
the biological point of view, and structurally unstable from
the mathematical point of view. It is a consequence of the mean
field approximation, and we expect such neutral cycling to be
broken under corrections to mean-field, or under the full
dynamics with finite range of interaction. This is indeed what
is seen in our numerical simulations for the full model with
three species: either the fixed point becomes attracting, or
the neutral cycles are replaced by a single attracting limit
cycle, with amplitude and period independent of the initial
conditions.

In addition to its non-robust prediction of neutral cycling of
the species, the mean-field approximation is not able to
explain our main numerical finding: that the fixed point
becomes stable for short-range interactions. From the
observations of Sect. \ref{sec:Spatial} and Figure
\ref{fig:ecosystems} it is likely that the stabilization of the
fixed point arises from the fact that the relative abundances
$x_i$ are macroscopic quantities that become averaged and
non-fluctuating when the microscopic structure contains many
different domains, as in Figure \ref{fig:ecosystems}a. Thus,
it is pertinent trying to extend the mean-field formalism to
describe the microscopic spatially-dependent configurations, as
done in the following section.

\subsection{Local mean-field and spatial stability}
\label{sec:localmf}

In this section we consider the species locations to be at the
nodes of a two-dimensional square lattice. Then the node index
$\nu$ can be considered to be a discrete two-dimensional vector
$\bnu$. For regular networks such as this one, the mean-field
approximation can be made local in space. This involves
removing correlations as
$\mean{n_{i,\mathbf{\bnu}}n_{i,\bnu}}\simeq\mean{n_{i,\bnu}}\mean{n_{i,\bnu}}$
while keeping the dependence of the mean quantities on the node
location.

Under this approximation, Eq.~\eqref{eq:ni} can be written as:
\begin{widetext}
\begin{equation}
 \frac{d}{ds}\rho_i(\bnu,s) =  \frac{1}{k(k-1)} \sum_{j} H_{ij}  \left[
  \left(\sum_{\blambda\in P_\bnu}\rho_i(\blambda,s)\right) \left(\sum_{\bmu\in P_\bnu}\rho_j(\bmu,s)\right)
  - \sum_{\blambda \in P_\bnu} \rho_i(\blambda,s) \rho_j(\blambda,s)
  \right]   -\frac{1}{2}\rho_i(\bnu,s).
\label{eq:localMF}
\end{equation}
\end{widetext}

We have used the notation $\mean{n_{i,\bnu}} \equiv
\rho_i(\bnu,s)$. Note that this equation reduces to Eq.
(\ref{eq:meanf}) when $\rho_i$ is homogeneous:
$\rho_i(\bnu,s)=x_i(s)$, $\forall \bnu$.

This new formulation allows us to assess the stability of
particular solutions against spatially-dependent perturbations.
For example we can focus on the stability of an homogeneous but
time-dependent solution $\rho_i(\bnu,s)=x_i(s)$ which verifies
Eq.~\eqref{eq:meanf}. To do so, we seek a solution to
Eq.~\eqref{eq:localMF} of the form
\begin{equation}
  \rho_i(\bnu,s)=x_i(s)+\delta_i(\bnu,s),
\end{equation}
and linearize to first order in $\delta$. With this,
Eq.~\eqref{eq:localMF} becomes
\begin{widetext}

\begin{eqnarray}
\frac{d\delta_i(\bnu,s)}{ds} &=& \sum_j \frac{H_{ij}}{k}
\left[ x_j(s) \sum_{\blambda\in P_\bnu} \delta_i(\blambda,s)
+  x_i(s) \sum_{\blambda\in P_\bnu} \delta_j(\blambda,s)\right]  -\frac12 \delta_i(\bnu,s)  \ .
\label{eq:localMFlin}
\end{eqnarray}
\end{widetext}

We introduce the Fourier transform of the perturbation:
$\hat\delta_i(\bq,s) = \sum_\bnu e^{i \bq\cdot\bnu}
\delta_i(\bnu,s)$, in terms of which Eq.
(\ref{eq:localMFlin}) reads:
\begin{eqnarray}
  \frac{d\hat\delta_i(\bq,s)}{ds} &=& \left[ -1 + 2 F(\bq) \sum_j H_{ij} x_j(s) \right]\hat\delta_i(\bq,s) \nonumber \\
                                &+& F(\bq) x_i(s) \sum_j H_{ij} \hat\delta_j(\bq,s) \ .
\label{eq:stability}
\end{eqnarray}
We have introduced the quantity
\begin{equation}
  \label{eq:fourarA}
  F(\bq)\equiv \frac{1}{k}\sum_{\blambda \in P_{\bm{0}}} e^{i\bq \cdot \blambda} \ ,
\end{equation}
which satisfies $F(\bq =\bm 0)=1$, $|F(\bq)|\le 1$, and
$F(\bq)\to 0$ as $|\bq|\to \infty$. Note that this quantity
contains information on the interaction range through the
dependence on $P_{\bm{0}}$ (i.e. through the set of neighbors of the origin). 

The simplest case to analyze is the stability of the interior
equilibrium point, i.e. $x_i(s)=x_i^*$, $\forall s$ as given by
Eq.~\eqref{eq:sts}. In this case Eq. \eqref{eq:stability} is a
linear system with constant coefficients, hence the stability
depends on the eigenvalues of the matrix of coefficients
$M_{ij}=F(\bq)x_i^* H_{ij} + [F(\bq)-1]\delta_{ij}/2$.  In
fact, because of Eq. (\ref{eq:sumdynamics}), there is always an
unstable eigenvalue $1/2$ for perturbations that bring the
dynamics out of the simplex. Thus, it is convenient to restrict
the dynamics to the simplex by using $\sum_{j=1}^g \delta_j=0$,
and then the matrix of the coefficients of Eq.
(\ref{eq:stability}) restricted to the first $g-1$ dimensions
is $M_{ij}=F(\bq)x_i^* (H_{ij}-H_{ig}) +
[F(\bq)-1]\delta_{ij}/2$, $i,j=1,...,g-1$.

For example, for $g=3$, the two eigenvalues of $M$ restricted
to the simplex can be explicitly calculated and read
\begin{equation}
  \lambda_\pm=-\frac{1-F}{2}
  \pm i \frac{F}{2}\sqrt{\frac{(2H_{12}-1)(2H_{13}-1)(2H_{23}-1)}{1 - 2(H_{12}-H_{13}+H_{23})}} \ .
\end{equation}
The argument of the square root is always positive when $H$
presents intransitive dominance cycles. Hence
\begin{equation}
  \textrm{Re}[\lambda_\pm]=-\frac{1-F(\mathbf q)}{2}\le 0
\end{equation}
and the equality holds if and only if $\mathbf q=\mathbf 0$.
This means that, within the mean-field approximation, the
steady and homogeneous solution $\rho_i(\bnu,s)=x_i^*$ is 
linearly stable against small spatial perturbations, except for
homogeneous perturbations, in which stability is marginal (a
fact that we already knew from the more general nonlinear
arguments in Sect. \ref{sec:mf}). Thus, the local mean-field
dynamics of Eq. (\ref{eq:localMF}) leads, for inhomogeneous
initial conditions close to the interior fixed point, to a
homogenization of the configuration, which then proceeds to
cycle neutrally around the fixed point. This is confirmed by
direct numerical simulation of Eq. (\ref{eq:localMF}). These
results hold for any value of the interaction range, contained
in $F(\bq)$. Thus, this local mean-field theory is not able to
explain the results from our stochastic model with structured
interactions. Namely, a transition from persistent inhomogeneous
configurations at short interaction range, which produce a
fully attracting fixed point for the macroscopic variable
$x_i(s)=\sum_\bnu \rho(\bnu,s)$, to a situation with
oscillatory dynamics that produces a repelling fixed point and
limit-cycle oscillations for $x_i(s)$ at large interaction
range.

Nevertheless, we can still use the local mean-field to gain
further insight into the dynamics, for example by analyzing the
stability with respect to inhomogeneous perturbations of a
homogeneous periodic solution $x_i(s)$ of Eq. (\ref{eq:meanf}).
In this case the stability equation (\ref{eq:stability}) is a
linear equation with periodic coefficients, which can be
analyzed with Floquet theory. The solutions can be written as a
linear combination of the functions \cite{gl94}
\begin{equation}
  f_i(s)e^{p_i s},\qquad i=1,\dots,g-1,
\end{equation}
where $f_i(s)$ are periodic (and hence bounded) functions of
time, with the same period $T$ as the functions $x_i(s)$, and
$p_i$ are the Floquet exponents given in terms of the
eigenvalues $\Lambda_i$ of the fundamental matrix $\Phi(s)$ of
system \eqref{eq:stability}, satisfying $\Phi(0)=I$, as
\begin{equation}
  \Lambda_i=e^{p_i T}.
\end{equation}
When all $p_i$ are negative, the perturbations decay and the
homogeneous solution $x_i(s)$ is recovered as time advances. We
have numerically evaluated $p_i$ for the case of three species,
$g=3$, and some values of the parameters of the system. For all
cases considered, $p_i$ has always negative real parts (except
for homogeneous perturbations, for which one finds neutral
stability), meaning that any initial inhomogeneous perturbation
tends to disappear. This agrees with direct simulation results
of Eq. (\ref{eq:localMF}). Thus, the long-term behavior of the local mean-field approach 
reduces to the standard homogeneous mean-field
treatment of Sect. \ref{sec:mf}. In contrast, simulation of the
stochastic model shows domains of the different species for
short interaction range.

However, the stability strength is not the same for all
parameter values. Let $M(s)$ be the matrix of time-dependent
coefficients of the system \eqref{eq:stability}. A necessary,
but not sufficient, condition for the homogeneous solutions
$x_i(s)$ to be unstable is that some eigenvalue of $M(s)$ has
positive real part for some time $s\in[0,T]$  (see a proof of a
similar result in \cite{klkh20}). During these times, even if
the trajectory turns out to be linearly stable, its stability
is reduced and more susceptible to non-infinitesimal
perturbations or noise. For the case of three species $g=3$,
and the interaction matrix $H$ given again by Eq.~(\ref{eq:H}), we have seen
that the matrix $M(s)$ has eigenvalues with positive real
parts, for some possible periodic trajectories $x_i(s)$,
provided $F(\bq) \gtrsim 0.67$. Since the maximum of $F(\bq)$
occurs at zero wavenumber and the width of this function
decreases with increasing $k$, the band of wavenumber
identified as `less-stable' shrinks as the interaction range,
quantified by $k$ increases. This is an indication (although
not a proof) that homogeneous periodic solutions would be more
robust for long-range interactions, and instabilities giving
rise to inhomogeneous configurations are more likely to occur
for short-range interactions. It is interesting to note that the
times at which the matrix $M(s)$ has more positive real-part of
eigenvalues coincide with the times at which some of the
components of the oscillatory solution $x_i(s)$ approach zero.

On general grounds, the local mean-field approximation should
represent some kind of coarse-graining of the original
stochastic system, and should be completed by noise terms to
gain accuracy. Under short-range interactions, appropriate
noise terms would be able to break the synchronization between
distant locations, and reproduce the domain structure
observed in the Monte Carlo simulations. However, we find
difficult to write analytical expressions for these noise terms
that would respect all the proper statistical constraints (for
example: reflect the multiplicative nature of birth-death
fluctuations, keep in time that $\sum_i \rho_i(\bnu,s)=1$,
etc.). Also, the complexity of such model would not be lower
than the original individual-based one. Thus, we have not
developed further this possibility. 

\section{\label{sec:app_toymodel} Effect of introducing correlations 
beyond mean-field}

In section \ref{sec:Spatial} we demonstrated that short-range interactions lead to the emergence of mono-specific clusters, effectively increasing intra-specific competition and 
stabilizing the dynamics.  As a further way to confirm that the decline of  inter-specific
competition is able to change the stability of the equilibrium,
making it stable for sufficiently reduced competition between
distinct species, we have studied a toy-model which shares
characteristics with our community model. It is built by
noticing that $\langle P_{ij} \rangle$ is just the time average
of the matrix $P_{ij}(s)$ in Eq. (\ref{eq:Ps}) of Appendix \ref{sec:theory}.
A way to correct the mean-field approximation $ P_{ij} \approx
x_i x_j$ is to introduce some correlations, $ P_{ij} \approx
c_{ij}x_i x_j$, making some ansatz for $c_{ij}$ and introducing it
into the exact equation (\ref{eq:xintermsofP}) (with
$\mean{x_i}\approx x_i$). We have explored the behavior of such
model in which correlations are implemented by
$c_{ij}=1-\epsilon$ if $i\neq j$ and $c_{ii}=1+\epsilon'$, with $\epsilon,\epsilon'>0$, 
resulting in an enhanced intra-specific competition with respect to
 inter-specific competition as $\epsilon$ and $\epsilon'$ are
increased. With this choice of $c_{ij}$ the resulting matrix
$P_{ij}$ does not have the proper statistical properties. In
particular the model does not respect that $\sum_i x_i=1$,
$\forall s$. However, this problem can be fixed by
constraining the dynamics onto the simplex by subtracting to
Eq. (\ref{eq:xintermsofP}), for each species $i$, the same term
$g^{-1}\sum_i G_i$, where $G_i$ is the right-hand-side of Eq.
(\ref{eq:xintermsofP}). It should be clear that this is not a
systematic approximation to our original system, but a toy
model useful to check the impact of varying the intra- and inter-specific competition balance. For example, for
$\epsilon'=0.01$ and the same dominance matrix used in the rest
of the paper, we have found that a Hopf bifurcation occurs at
$\epsilon = \epsilon_c\approx 0.01975$, so that relative
species abundances undergo limit cycle oscillations for
$\epsilon<\epsilon_c$ but the fixed point becomes stable and
attracting when the inter-specific competition is further
reduced, $\epsilon>\epsilon_c$. These are the same type of
states and the same transition that is encountered in our
stochastic model when decreasing the interaction range.

%

\end{document}